\documentclass[conference]{IEEEtran}

\usepackage{titlesec}
\usepackage{color,soul}
\usepackage{graphicx}
\usepackage{subcaption}
\usepackage{enumitem}

\usepackage[acronym,toc,shortcuts]{glossaries}
\usepackage{multirow}
\makeglossaries
\newacronym{MANO}{MANO}{Management and Orchestration}
\newacronym{OAM}{OAM}{Operations, Administration, and Maintenance}

\title{Enhancing Cloud-Native  Resource Allocation with Probabilistic Forecasting Techniques in O-RAN}

\author{
\IEEEauthorblockN{
Vaishnavi Kasuluru\IEEEauthorrefmark{1}, Luis Blanco\IEEEauthorrefmark{1}, Engin Zeydan\IEEEauthorrefmark{1}, Albert Bel\IEEEauthorrefmark{1} Angelos Antonopoulos\IEEEauthorrefmark{2} 
}

\IEEEauthorblockA{\IEEEauthorrefmark{1}Centre Tecnològic de Telecomunicacions de Catalunya (CTTC), Barcelona, 08860 Spain.}
\IEEEauthorblockA{\IEEEauthorrefmark{2}Nearby Computing S.L., Barcelona, 08006 Spain.}
\IEEEauthorblockA{Email: 
\IEEEauthorrefmark{1}\{vkasuluru, lblanco, ezeydan, abel\}@cttc.es,
\IEEEauthorrefmark{2}aantonopoulos@nearbycomputing.com
}
}
\makeglossaries
\newacronym{3GPP}{3GPP}{The 3rd Generation Partnership Project }
\newacronym{5G}{5G}{Fifth Generation}
\newacronym{AAA}{AAA}{Authentication, Authorization and Accounting}
\newacronym{ARIMA}{ARIMA}{AutoRegressive Integrated Moving Average}
\newacronym{CPU}{CPU}{Central Processing Unit}
\newacronym{GRU}{GRU}{Gated Recurrent Unit}
\newacronym{O-RAN}{O-RAN}{Open Radio Access Network}
\newacronym{QoS}{QoS}{Quality of Service}
\newacronym{IoT}{IoT}{Internet of Things}
\newacronym{QoE}{QoE}{Quality of Experience}
\newacronym{URLLC}{URLLC}{Ultra-Reliable and Low Latency Communication}
\newacronym{mMTC}{mMTC}{massive Machine-Type Communication}
\newacronym{PRB}{PRB}{Physical Resource Block}
\newacronym{rApp}{rApp}{radio App}
\newacronym{RAN}{RAN}{Radio Access Network}
\newacronym{RIC}{RIC}{RAN Intelligent Controller}
\newacronym{xApp}{xApp}{eXtended application}
\newacronym{SFF}{SFF}{Simple-Feed-Forward}
\newacronym{LSTM}{LSTM}{Long-Short Term Memory}
\newacronym{SN}{SN}{Seasonal-Naive}
\newacronym{MLP}{MLP}{ Multi-Layer Perceptron}
\newacronym{RNN}{RNN}{Recurrent Neural Network}
\newacronym{MSE}{MSE}{Mean Square Error}
\newacronym{MAE}{MAE}{Mean Absolute Error}
\newacronym{MAPE}{MAPE}{Mean Absolute Percentage Error}
\newacronym{MASE}{MASE}{Mean Absolute Scaled Error}
\newacronym{ND}{ND}{Normalized Deviation}
\newacronym{QL}{QL}{Quantile Loss}
\newacronym{AI}{AI}{Artificial Intelligence}
\newacronym{ML}{ML}{Machine Learning}
\newacronym{SLA}{SLA}{Service Level Agreement}
\newacronym{DNN}{DNN}{Deep Neural Network}
\newacronym{DYNp}{DYNp}{Dynamic Percentile Adjustment Approach}
\newacronym{ECDF}{ECDF}{Empirical Cumulative Distribution Function}
\newacronym{DeepAR}{DeepAR}{Deep Autoregressive Recurrent network}
\newacronym{CI/CD}{CI/CD}{Continuous Integration / Continuous Delivery or Continuous Deployment}
\newacronym{CAPEX}{CAPEX}{Capital Expenditure}
\newacronym{OPEX}{OPEX}{Operating Expense}
\newacronym{O-RU}{O-RU}{Open-Radio Unit}
\newacronym{O-DU}{O-DU}{Open-Distribution Unit}
\newacronym{O-CU}{O-CU}{Open-Central Unit}

\usepackage{fancyhdr}
\usepackage{geometry}
\usepackage{lipsum} 

\begin{document}

\maketitle

\begin{abstract}

The need for intelligent and efficient resource provisioning for the productive management of resources in real-world scenarios is growing with the evolution of telecommunications towards the 6G era. Technologies such as \ac{O-RAN} can help to build interoperable solutions for the management of complex systems. Probabilistic forecasting, in contrast to deterministic single-point estimators, can offer a different approach to resource allocation by quantifying the uncertainty of the generated predictions.  This paper examines the cloud-native aspects of \ac{O-RAN} together with the {\ac{rApp}} deployment options. The integration of probabilistic forecasting techniques as a {\ac{rApp}} in {O-RAN} is also emphasized, along with case studies of real-world applications.  Through a comparative analysis of forecasting models using the error metric, we show the advantages of {\ac{DeepAR}} over other deterministic probabilistic estimators. Furthermore, the simplicity of {\ac{SFF}} leads to a fast runtime but does not capture the temporal dependencies of the input data. Finally, we present some aspects related to the practical applicability of cloud-native \ac{O-RAN} with probabilistic forecasting.







\end{abstract}


\section{Introduction} 


The conventional \ac{RAN} architecture used in mobile networks used to have considerable problems with network scalability based on demand, insufficient flexibility, and tightly coupled vendor-specific hardware and software configurations. These limitations were addressed together with the virtualization and disaggregation of \ac{RAN} network components, resulting in the evolution towards \ac{O-RAN}. \ac{O-RAN} aims to achieve openness, intelligent decision making, and simplicity in the integration of third-party services \cite{polese2023understanding}. Nevertheless, network optimization and resource allocation remain a major challenge, especially as network management becomes more complex. In addition, real-time decision making is a major challenge due to fluctuating network requirements. Cloud-native \ac{O-RAN}s have significantly changed the design, management, and operation of the network component. Containerization, dynamic orchestration, and the management of microservices in cloud-native \ac{O-RAN} have helped to create an effective, scalable, and flexible network \cite{song2023micro}. However, the complexity of resource allocation has increased now, requiring more sophisticated techniques \cite{KAZEMIFARD2021107809}.

In modern telecommunications, effective resource allocation is crucial to meet dynamic network requirements. With the evolution of 6G technologies, the requirements for computing power, bandwidth, storage capacity, and orchestration of resources have become more specific depending on the application and service. 
Radio and computing resources must be allocated dynamically to cope with variable network traffic. In addition, latency- and reliability-dependent \ac{URLLC} applications, the high bandwidth requirements of \ac{mMTC} applications, and the growing number of \ac{IoT} devices must also be managed. Effective management of resources can lead to the optimization of operating costs and energy consumption; at the same time, the network can provide better \ac{QoS} and \ac{QoE} to end users \cite{article}. 

In wireless networks, \ac{PRB}s are considered the most important resource as they play a crucial role in optimal radio spectrum utilization. 
The landscape of existing resource allocation techniques in \ac{O-RAN} is diverse and reflects the ongoing evolution of mobile network architectures. Several studies have delved into conventional approaches to resource allocation, highlighting the challenges posed by dynamic and distributed radio access networks \cite{xu2021survey, sharma2021resource}. However, there is a lack of resource allocation with probabilistic forecasts that would enable decision-making in the \ac{O-RAN}.

Advanced and efficient network management can be achieved by incorporating probabilistic forecasting into resource provisioning applications of \ac{O-RAN} \cite{9690624}. Traditional deterministic single-point forecasting models fail to address modern telecommunication networks' dynamic nature and uncertainty. Various time series analyses and machine learning algorithms have been used in previous works to model and predict network parameters \cite{articlemdpi}. However, probabilistic forecasting models offer a spectrum of possible outcomes along with their uncertainty in terms of probability. Using this approach and understanding the probabilities associated with different scenarios, network operators can make knowledgeable decisions regarding resource provisioning in real-time to optimize network performance and deliver services effectively 
. Furthermore, it can adapt to the forecasted traffic pattern, service demand, and spectrum requirement, allowing for proactively adjusting the distribution of resources based on predictions \cite{9690624}. Probabilistic forecasting can also find emerging trends and handle data non-stationarity efficiently, making it more suitable for user-centric \ac{O-RAN} infrastructure \cite{10266607}. Such forecasting is beneficial in a cloud-native environment to optimize resource provisioning and reduce resource wastage through effective prediction regardless of dynamic network demands.

\subsection{Transition to Cloud Native Architectures}

The cloud-native O-RAN architecture aims to enable better network scalability, agility, and effective management and operations through key changes in network management and deployment strategies. Compared to traditional architectures, the use of containers, \ac{CI/CD}, and microservices in cloud-native technology applications contributes to faster deployment of services, simplifies the integration of third-party applications, and improves fault tolerance. Cloud-native deployment is in line with the O-RAN vision of an open, intelligent, and flexible network and paves the way for the development and deployment of effective resource management and provisioning strategies to meet real-time network requirements\cite{song2023micro}. The main drawback is that the dynamic nature of the network presents challenges in resource management and allocation. Advanced resource provisioning solutions and efficient orchestration of resources are essential to deal with the scalable and volatile nature of containers in a multi-vendor infrastructure. Cloud-native O-RAN ensures a flexible and efficient network but also demonstrates the need for innovative techniques to manage the complexity of an advanced network infrastructure.

The advent of cloud-native paradigms has also significantly reshaped resource management strategies, particularly in the context of \ac{O-RAN} \cite{song2023micro, liu2022cloud, mohammadi2023athena}. The authors in \cite{mohammadi2023athena} propose a new generation of \ac{MANO}/\ac{OAM} that follows the principles of cloud-native. 



\subsection{Contributions}

Significant contributions are made in this paper within the scope of AI-enabled resource allocation in cloud-native \ac{O-RAN} environments. Initially, \ac{O-RAN} architecture equipped with containerizing and integrating resource provisioning \ac{rApp} in \ac{O-RAN} is briefed in section 2. Then, the probabilistic forecasting methods used for accurate prediction of resources along with containerization of the resource provisioning \ac{rApp} are explained in section 3. Moreover, the metrics used to highlight the efficiency of \ac{rApp} in terms of error, CPU, and memory usage are emphasized in section 4. Here, the effect of data length on the prediction and comparison of the performance of different estimators is also briefed. Finally, some case studies on cloud-native O-RAN in real-world applications are provided in section 5, followed by a conclusion in section 6.

\section{\ac{O-RAN} Architecture}

\begin{figure}[htp!]
\centering
\includegraphics[width=0.8\linewidth]{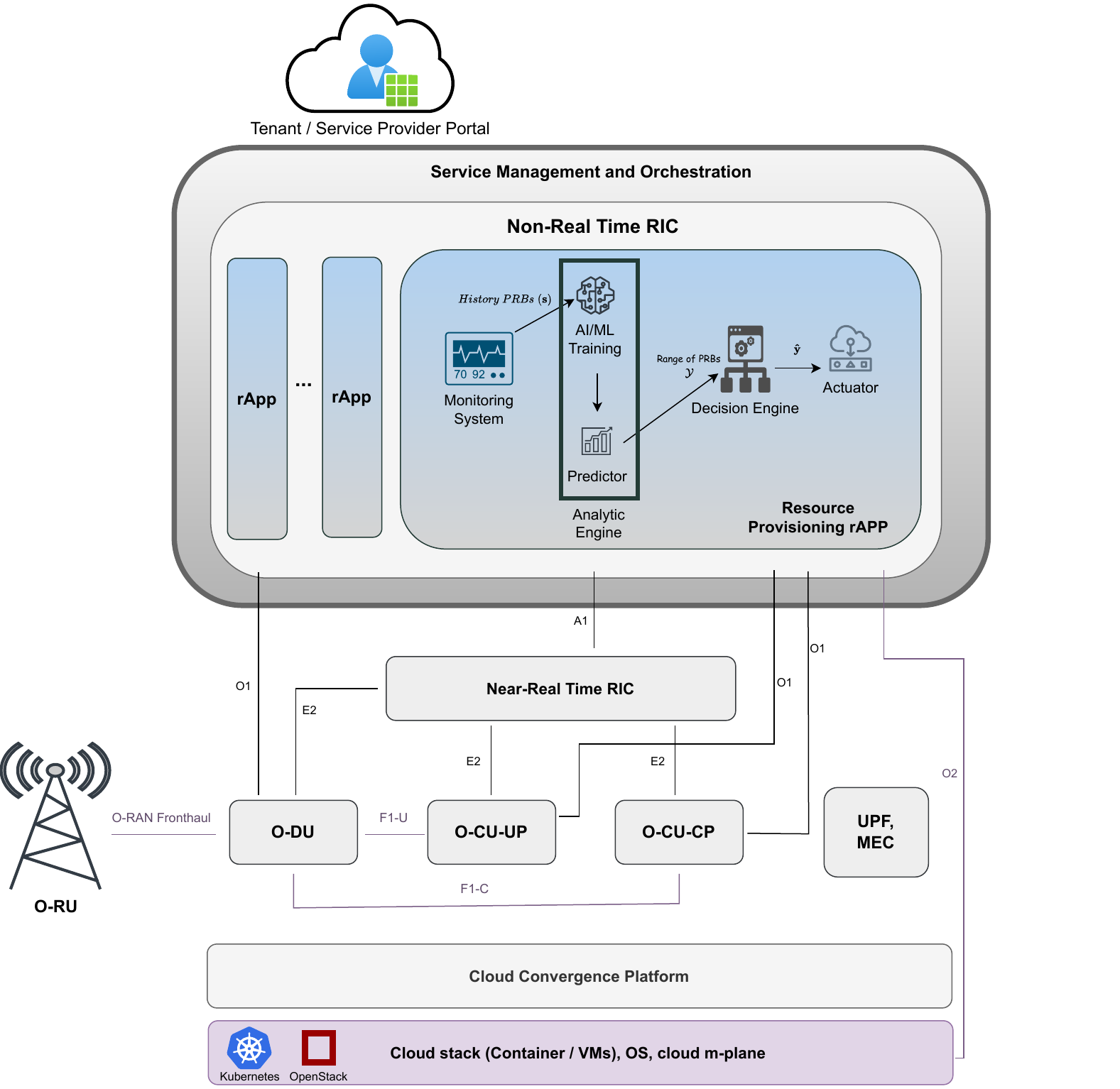}
\caption{\ac{O-RAN} architecture with probabilistic forecast as \ac{rApp}.}
\label{O-RAN}
\end{figure}

The traditional RAN is highly dependent on hardware components, leading to vendor lock-in issues. It is causing a huge rise in \ac{CAPEX} and \ac{OPEX} costs. It becomes very challenging for network providers to integrate intelligence and build a collaborative and reliable network. Therefore, it's crucial to establish next-generation RAN solutions with global, self-reliant hardware and software-defined technology that are independent of vendors. Virtualization and RAN dis-aggregation are the key technologies for the concept of O-RAN, whose main pillars are openness and intelligent resource management\cite{polese2023understanding}. Fig. \ref{O-RAN} shows the Open RAN architecture with probabilistic forecasting based resource provisioning as \ac{rApp}. The main components of O-RAN are: 

\begin{itemize}[leftmargin=1.5mm]
    \item \ac{O-RU}, \ac{O-DU} and \ac{O-CU} whose functionalities are similar to that in 5G dis-aggregated RAN except with added support of \ac{O-RAN} based specifications and interface.
    \item Near-Real Time \ac{RIC}  to control/optimize RAN elements and resources based on fine-grained data using online \ac{AI}/\ac{ML} based services. It is suitable for applications with latency requirements between 10ms and 1s.
    \item Non-Real Time \ac{RIC} to control/optimize RAN elements and resources based on coarse-grained data using online \ac{AI}/\ac{ML} services. It is suitable for applications with latency requirements greater than 1s. It also provides policy-based guidance to near-real Time \ac{RIC}. In our work, the containerized resource provisioning \ac{rApp} is placed in Non-Real Time \ac{RIC}. It consists of 3 main components, namely,
    \begin{itemize}
        \item Monitoring System which receives required information from Tenants (in our case, History \ac{PRB}s $\mathbf{s}$) and forwards it to other elements in rApp. 
        \item Analytical Engine is responsible for data preprocessing, train-test split, model training, and prediction using probabilistic and deterministic estimators.
        \item Decision Engine receives the range of estimated PRBs $\mathbf{\mathcal{Y}}$ with the corresponding associated probabilities from the predictor and applies decision making logic to obtain the exact number of \ac{PRB}s $\mathbf{\hat{y}}$ to be allocated to the tenant in next time instances.
        \item Actuator passes the information on a number of \ac{PRB}s to be allocated to the \ac{O-DU} via the O1 interface.
    \end{itemize}
    \item Containerized/virtualized \ac{RIC} components are deployed at some point between the cell site and the core network. They can be placed either on the edge or in the regional cloud network, depending upon the usage scenario.
\end{itemize}

\section{Methodology}

\subsection{Overview of Cloud Native Resource Allocation in O-RAN}


Cloud-native principles are increasingly being integrated into O-RAN to improve resource allocation strategies and optimize network performance. Cloud-native resource allocation in O-RAN includes the use of containerized applications, microservices, and dynamic orchestration for efficient resource utilization. 
Applications are encapsulated in lightweight containers, enabling portability and consistency across environments. \ac{O-RAN} also uses a microservices architecture that splits monolithic applications into smaller, independently deployable services. Each microservice focuses on a specific function and facilitates scalability and agility in resource allocation. Dynamic orchestration frameworks such as Kubernetes play a central role in the allocation of cloud-native resources. They automate the deployment, scaling and management of containerized applications and ensure optimal resource utilization based on real-time demand. Service mesh technologies such as Istio\footnote{Available: https://istio.io/, Online: January-2023.} improve communication and control between microservices. They provide features such as load balancing, traffic management, and resilience and contribute to efficient resource allocation in a distributed and scalable manner. Resource allocation strategies such as autoscaling, resource pooling, or declarative resource management techniques (e.g., using Infrastructure as Code (IaC) or configuration files) can be used to take advantage of scalability, flexibility, operational efficiency, and cost optimization features of cloud-native deployments.

\subsection{Introduction to Probabilistic Forecasting Techniques}
Probabilistic estimators like \ac{SFF}, \ac{DeepAR}, and Transformer are being used in a wide variety of applications as they are more advantageous when compared to single-point forecast models like \ac{LSTM} and \ac{GRU}s. They are gaining more attention due to their ability to provide a range of possible outcomes along with information about their uncertainty of occurrence.

\textbf{\ac{SFF}} forecasting works based on a simple feed-forward neural network. Neural networks, also called Multi-Layer Perceptrons (MLP), are built with a combination of an input layer, a hidden layer, and an output layer. The number of layers and neutrons rises with the task complexity. Here, the information flows in the forward direction from the input to the output layer from neurons. they don’t have any feedback loop. In the training phase, the model predicts possible outcomes using the initial assigned weights. Furthermore, the actual and predicted values are compared to adjust the network weights in each layer to improve the predictions. This parameter adjustment is called Backpropagation. In the prediction phase, the trained model parameters are used to process the new information and predict the appropriate possible outcomes along with their probability of occurrence in the form of uncertainty. 

\textbf{\ac{DeepAR}} forecasting was developed by Amazon \cite{SALINAS20201181}. They work based on the Recurrent Neural Network (RNN) framework. They are autoregressive recurrent network encoder-decoder that uses an encoder-decoder architecture with a sequence-to-sequence model based on \ac{LSTM} cells. The \ac{DeepAR} model is trained to maximize the likelihood function. they use a negative log-likelihood loss function to optimize the neural network parameters. During prediction, the learned likelihood function is used to forecast the possible output. They have an exceptional understanding of patterns and relationships in historical information. The \ac{DeepAR} accepts inputs as time series information and collects the temporal information using RNN. The RNN predicts future outcomes using historical data and improves the accuracy of prediction through covariates. The obtained output of RNN passes through the connected layers to generate a probabilistic forecast of future outcomes. \ac{DeepAR} is more versatile in capturing regular and irregular trends in time series data and more robust towards data seasonality. Since the mean and variance of the distribution are considered in the prediction of the loss function, the model performance is more accurate and accounts for the uncertainty in the forecasting.

\textbf{Transformers} were majorly used in natural language processing. They slowly gained popularity in time-series forecasting because of their proficient handling of long-term dependencies and finding complex trends in input data. Later, their architecture was improved to make them suitable for probabilistic distribution approaches \cite{NIPS2017_3f5ee243}. The transformer architecture consists of serially structured layers of encoders and decoders. The encoder layers process the input data, while the decoder layers generate the output prediction sequences. A self-attention mechanism is used to find each major time stamp. It assists the models to focus on specific parts of the inputs that have a major impact on the ongoing predictions. The transformer provides future predictions along with their probability distribution. This distribution can be further split to obtain a range of possible outcomes and their probability of occurrence.

To summarize, the \ac{SFF} estimators are less complex and are very well suited for small datasets, as they are easy to implement and can recognize complex non-relationships well. \ac{SFF} estimator cannot handle temporal dependencies that are crucial for time-series forecasting, which is a significant disadvantage. \ac{DeepAR} and Transformer overcome this challenge because of their robustness and versatility. \ac{DeepAR} and Transformers can be used for a wide range of applications like decision-making, admission control, and resource allocation in telecommunications, especially for tasks based on probabilistic forecasts. Compared to deterministic forecast models, probabilistic approaches such as \ac{DeepAR} and Transformer can outperform traditional \ac{LSTM} predictions regarding performance, reliability, and accuracy, helping make more informed decisions. They can capture uncertainties well, which is crucial in applications with uncertain or variable data. 


\subsection{Integration Approach of Probabilistic Forecasting as Radio Application (\ac{rApp}) in \ac{O-RAN}}

In the \ac{O-RAN} context, the integration of probabilistic forecasting within the Radio Application (\ac{rApp}) framework is a strategic approach aimed at optimizing wireless communication systems. This integration involves using \ac{RAN} intelligence, a crucial element, where the network dynamically manages radio resources, making real-time decisions to adapt to diverse conditions and user requirements. To formalize this approach, the \ac{O-RAN} Alliance has introduced a framework incorporating both Non-Real and near-real Time RAN Intelligent Controllers (RICs) and associated interfaces (E2, O1, and O2). By combining the Service Management and Orchestration (SMO) capabilities with the Non-Real Time \ac{RIC} and \ac{rApp}s, a holistic perspective of the entire O-cloud components and the available network services is achieved. This integrated component empowers the creation of high-level policies and facilitates the lifecycle management of network services with a time granularity exceeding 1s, enhancing the overall efficiency and adaptability of the O-RAN ecosystem.



\section{Experimental Setup and Results}

\subsection{Experimental Setup}
The performance of deterministic and probabilistic forecast estimators is shown in this section. Python programming was used along with the Gluonts library \cite{alexandrov2018gluonts} to analyze the resource provision rApp. Three probabilistic forecasting algorithms, namely \ac{SFF}, \ac{DeepAR}, and Transformer, were used to predict the DL \ac{PRB}s required for next 24-hours based on historical data. The historical \ac{PRB} data of the tenants can be obtained from the \ac{O-DU} in the \ac{O-RAN} architecture via the O1 interface. The dataset was created by simulating traffic and mobility patterns for a different number of end users \cite{9933014}. Initially, the history data is obtained from the tenant for training the model, and their prediction is performed to evaluate the accuracy of each estimator. The history \ac{PRB} data $\mathbf{s}$ is divided into training and test data in a ratio of 80:20. The training data is used to adjust the machine learning parameters and train the model for each forecasting algorithm. In our work, training data sets of different lengths namely 2 weeks, 4 weeks, 10 weeks and 20 weeks are used. The test data $\mathbf{x}$, on the other hand, is used to evaluate and compare the performance of each model. The prediction length is fixed to 24 hours, irrespective of changing training data length. Furthermore, the hyperparameters considered for estimators are as follows: \ac{SFF}: epocs=5, batch size=1, hidden layer dimension=[40,40], and number of evaluation samples=100. DeepAR: epocs=5, batch size=1, \ac{RNN} Layers=2, number of cells per \ac{RNN}=40, and number of evaluation samples=100. Transformer: epocs=5, batch size=1,  number of evaluation samples=100, dimension of transformer network=32, inner-hidden layers of transformer's feedforward network dimension=4, and context length=24. LSTM: Sequential model, epocs=5, batch size=1, neurons=1, and optimizer: adam.
The performance evaluation is done in the following ways: Initially, using error metrics like \ac{MSE}, followed by calculating training time and prediction time, and finally by monitoring the CPU and memory usage during training and prediction. 

In time series forecasting, error information can be obtained by evaluation metrics, where the array of forecasted probability distribution $\mathbf{y}$ are compared with the actual test instances $\mathbf{x}$. Here, the \ac{MSE} metric is used to analyze the prediction error. \ac{MSE} Measures the average of the squared differences between the forecasted and actual values. The output of probabilistic forecasting estimators is the range of \ac{PRB}s ($\mathbf{\mathcal{Y}}$). It is a combination of $[\mathbf{y}_1,\mathbf{y}_2...\mathbf{y}_{\mathcal{N}}]$, where $\mathcal{N}$ is the length of test instance and each $\mathbf{y}_i$ is a column vector that looks like $[y_{i1},y_{i2},...y_{in}]^T$ with $n$ being the percentiles ranging from 1-th to 99-th. These percentiles tell us about its probability of occurrence. The input test data $\mathbf{x}$ equals $[x_1,x_2,...x_{\mathcal{N}}]$. The \ac{MSE} is the most commonly
used metric for point forecasting. For probability forecasting, it is calculated by initially taking the mean of the $\mathbf{\mathcal{Y}}$ at every test time instance to obtain $\mathbf{\hat{y}}$. Here, $\mathbf{\hat{y}} = mean(\mathbf{\mathcal{Y}})$). The $\mathbf{\hat{y}}$ is now just an array containing single-point forecast values corresponding to each time instance i.e., $[\hat{y}_1,\hat{y}_2,...\hat{y}_{\mathcal{N}}]$. The \ac{MSE} is formulated as,

\begin{equation}
MSE = \frac{1}{\mathcal{N}} \sum_{i=1}^{\mathcal{N}} |x_i-\hat{y}_i|^2
\end{equation}
\noindent 





\subsection{Results and Analysis}




Table I compares probabilistic models like \ac{SFF}, \ac{DeepAR}, and Transformer, with deterministic \ac{LSTM} as the base model for different data lengths. The main aim here is to analyze the impact of data size on the error metric \ac{MSE}, model training time, and future prediction time. The \ac{MSE} is calculated using Equation 1. The size of data lengths considered here are 2 weeks, 4 weeks, 10 weeks, and 20 weeks. From the \ac{MSE} values, it can observed that the model performance is bad when trained with the small data set. The \ac{MSE} is low, around 51.20, 1.38, 0.03, and 0.09 for \ac{LSTM}, \ac{SFF}, \ac{DeepAR}, and Transformer, respectively, with 20 weeks of data. When compared among the estimators, \ac{LSTM}s performance is the worst, irrespective of the data length, as the deterministic base models are more sensitive to outliers, uncertainties, and data non-stationarity. Among the probabilistic estimators, \ac{SFF} fails to perform better and has high uncertainty because it can not handle the temporal dependencies effectively.

SFF takes less time to train when compared to other estimators due to the low model complexity. As discussed earlier, SFF networks are a simple feed-forward neural network. \ac{LSTM}, on the other hand, takes longer training time and increases more with the data length as they use a gating mechanism to run RNNs in a sequential form and control the information flow. The number of parameters required is quite high, affecting the computation complexity during training. The transformer with a data length of 20 weeks takes the longest time to train as the self-attention mechanism scales quadratically with the input data length. Regarding prediction time, the test data length is fixed to 24 hours, irrespective of the training data length. Hence, the prediction time of probabilistic estimators is not affected by changes in data length. The values recorded in the table are the average of the prediction time of the dataset with different sizes. The |\ac{LSTM} takes a longer time to predict due to sequential data processing with a negligible impact of data length on prediction time, ranging from 1193ms. 

\begin{table}[ht]
\centering
\caption{Comparisons of the performance of different forecast methods under different times.}
\begin{tabular}{|l|llll|}
\hline
\multicolumn{1}{|c|}{\multirow{2}{*}{\textbf{\begin{tabular}[c]{@{}c@{}}Forecast \\ Models\end{tabular}}}} & \multicolumn{4}{c|}{\textbf{Training Dataset Length}} \\ \cline{2-5} 
\multicolumn{1}{|c|}{} & \multicolumn{1}{c|}{\textbf{2 Weeks}} & \multicolumn{1}{c|}{\textbf{4 Weeks}} & \multicolumn{1}{c|}{\textbf{10 Weeks}} & \textbf{20 Weeks} \\ \hline
 & \multicolumn{4}{c|}{\textbf{MSE}} \\ \hline
\textbf{LSTM} & \multicolumn{1}{l|}{59.7165} & \multicolumn{1}{l|}{58.170} & \multicolumn{1}{l|}{53.055} & \textbf{51.20} \\ \hline
\textbf{SFF} & \multicolumn{1}{l|}{6.75} & \multicolumn{1}{l|}{2.47} & \multicolumn{1}{l|}{1.80} & \textbf{1.381} \\ \hline
\textbf{DeepAR} & \multicolumn{1}{l|}{5.85} & \multicolumn{1}{l|}{4.99} & \multicolumn{1}{l|}{0.006} & \textbf{0.003} \\ \hline
\textbf{Transformer} & \multicolumn{1}{l|}{9.25} & \multicolumn{1}{l|}{3.66} & \multicolumn{1}{l|}{0.673} & \textbf{0.009} \\ \hline
 & \multicolumn{4}{c|}{\textbf{Training time (Seconds)}} \\ \hline
\textbf{LSTM} & \multicolumn{1}{l|}{66.992} & \multicolumn{1}{l|}{74.371} & \multicolumn{1}{l|}{76.915} & 79.516 \\ \hline
\textbf{SFF} & \multicolumn{1}{l|}{5.616} & \multicolumn{1}{l|}{5.995} & \multicolumn{1}{l|}{6.119} & 6.3 \\ \hline
\textbf{DeepAR} & \multicolumn{1}{l|}{24.144} & \multicolumn{1}{l|}{25.781} & \multicolumn{1}{l|}{31.504} & 48.057 \\ \hline
\textbf{Transformer} & \multicolumn{1}{l|}{24.355} & \multicolumn{1}{l|}{29.517} & \multicolumn{1}{l|}{34.721} & 85.21 \\ \hline
 & \multicolumn{4}{c|}{\textbf{Prediction time (Milliseconds)}} \\ \hline
\textbf{LSTM} & \multicolumn{4}{c|}{1193} \\ \hline
\textbf{SFF} & \multicolumn{4}{c|}{0.26} \\ \hline
\textbf{DeepAR} & \multicolumn{4}{c|}{0.31} \\ \hline
\textbf{Transformer} & \multicolumn{4}{c|}{0.29} \\ \hline
\end{tabular}
\end{table}

For the calculation of \ac{CPU} and memory usage during training and prediction, a data length of 20 weeks is used. the x-axis in all figures represents the run time in seconds. During training, the run time includes time to perform data preprocessing, train-test split, train the model, and model storage. On the other hand, during prediction, run time includes time to get test data, get the trained model, perform prediction, and post-processing. Here, 2 \ac{CPU} cores are used while performing analysis. 

\begin{figure*}[htp!]
    \centering
    \begin{subfigure}[b]{.48\linewidth}
        \centering
\includegraphics[width=\linewidth]{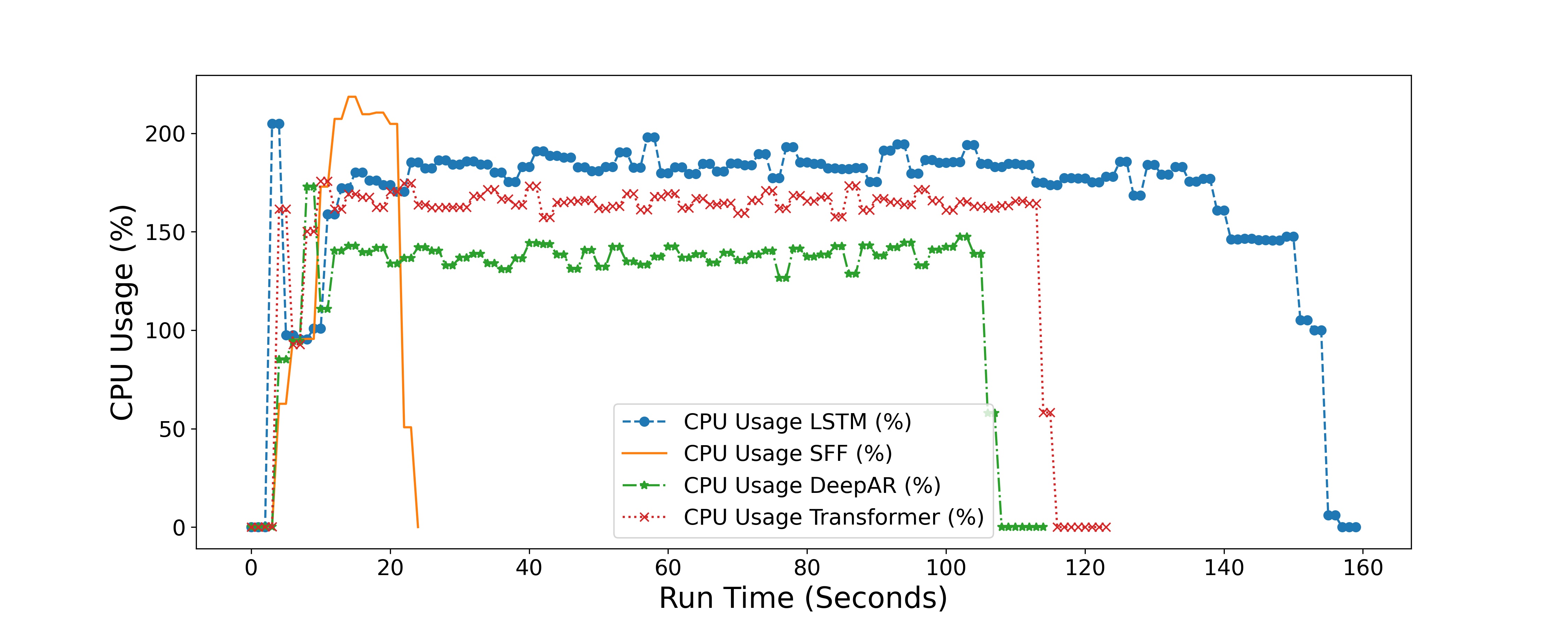}
        \caption{}
        \label{cpu_usage_plot}
    \end{subfigure}%
    \begin{subfigure}[b]{.48\linewidth}
        \centering
\includegraphics[width=\linewidth]{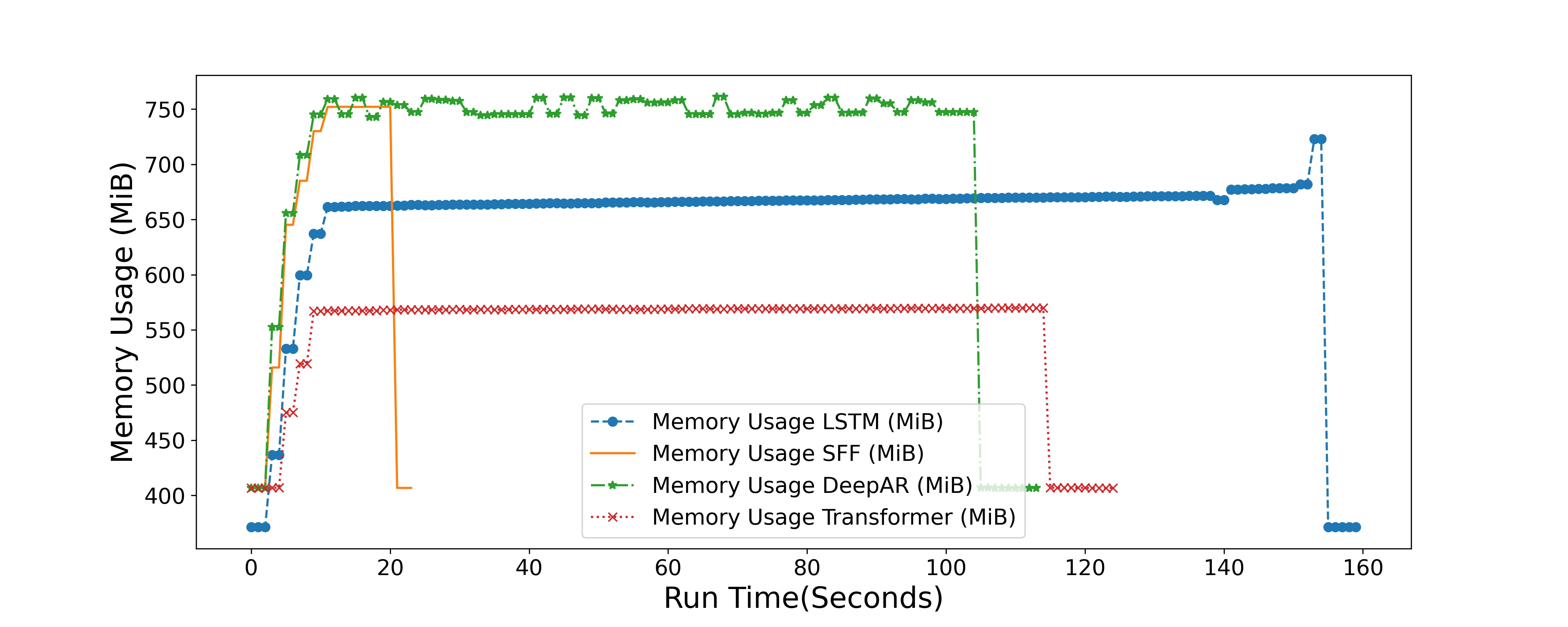}
        \caption{}
        \label{memory_usage_plot}
    \end{subfigure}
    \caption{Training (a) CPU Usage , (b) Memory Usage}
    \label{cpu_memory_usage_plot}
\end{figure*}

Figure \ref{cpu_memory_usage_plot} shows the \ac{CPU} and memory required during the training of \ac{LSTM}, \ac{SFF}, \ac{DeepAR}, and Transformer. Figure \ref{cpu_usage_plot} shows the line graph of CPU usage during model training with the x-axis being the run time in seconds and the y-axis being the \ac{CPU} Usage in percentage \%. \ac{LSTM} has a high \ac{CPU} usage of 200\% for a longer duration of around 160 seconds due to complex calculations involved during pre-processing and also during training along with sequential information processing. SFF has \ac{CPU} usage of 200\% for very little time, around 25 seconds, due to its simple model architecture. \ac{DeepAR} and Transformer have moderate CPU usage due to complex calculations and self-attention mechanisms, respectively. Similarly, for memory usage in figure \ref{memory_usage_plot}, The \ac{LSTM} uses a high amount of memory for a longer duration to maintain multiple state vectors. \ac{DeepAR} memory usage is 750MiB, which is higher than \ac{LSTM} but for a shorter time as both use autoregressive recurrent structures. The transformer requires the lowest memory of around 550MiB for around 120 seconds, as there are many parameters along with complex self-attention mechanisms.

\begin{figure*}[htp!]
    \centering
    \begin{subfigure}[b]{.48\linewidth}
        \centering
\includegraphics[width=\linewidth]{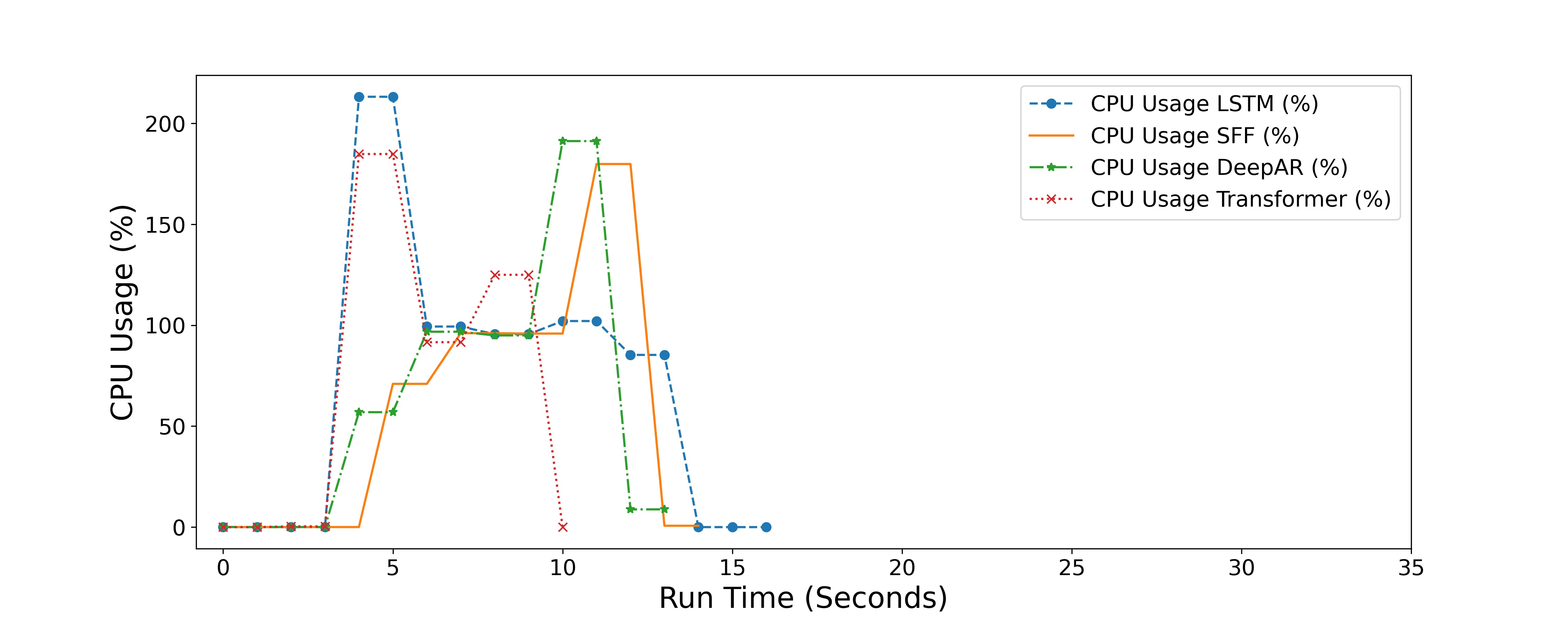}
        \caption{}
        \label{cpu_usage_plot_test}
    \end{subfigure}%
    \begin{subfigure}[b]{.48\linewidth}
        \centering
\includegraphics[width=\linewidth]{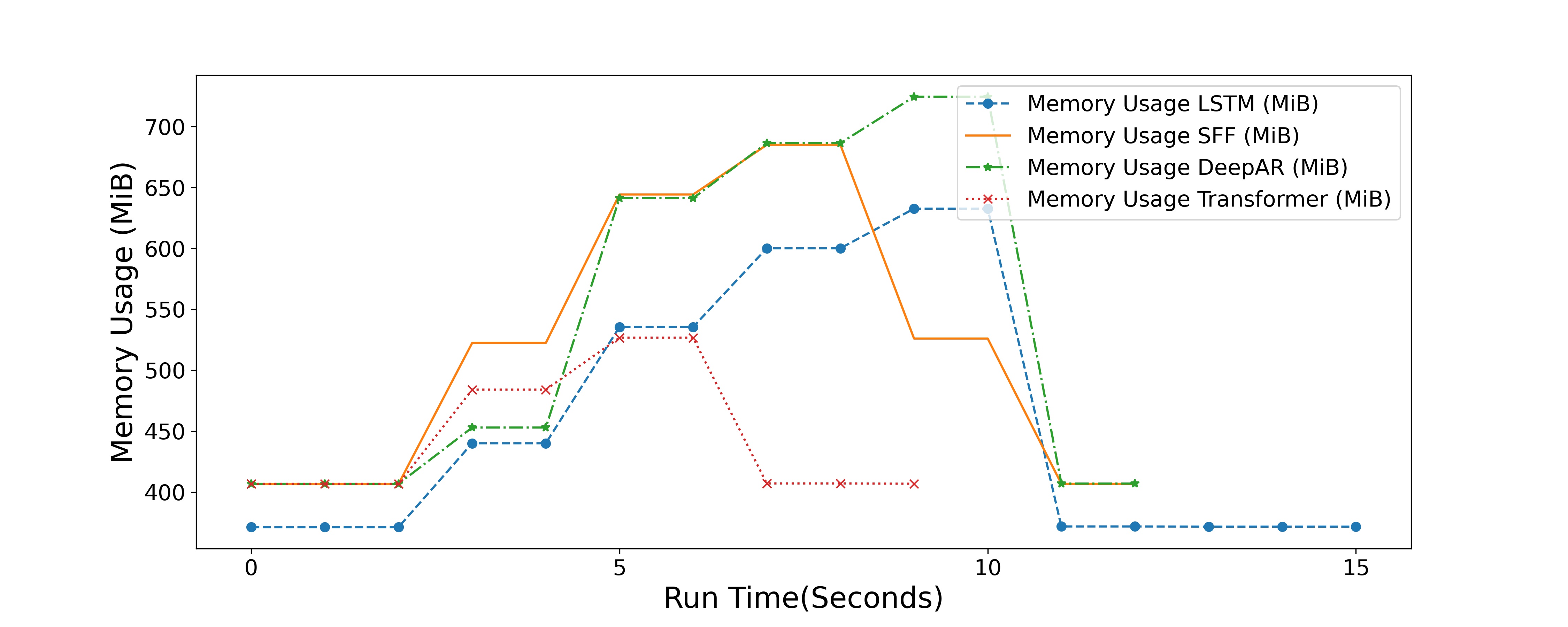}
        \caption{}
        \label{memory_usage_plot_test}
    \end{subfigure}
    \caption{Prediction (a) CPU Usage , (b) Memory Usage}
    \label{cpu_memory_usage_plot_test}
\end{figure*}

Figure \ref{cpu_memory_usage_plot_test} shows the \ac{CPU} and memory usage of \ac{LSTM}, \ac{SFF}, \ac{DeepAR}, and Transformer during prediction. From Figures \ref{cpu_usage_plot_test} and \ref{memory_usage_plot_test}, it can be understood that the transformer requires low \ac{CPU} and memory usage compared to other estimators with the least time of 10 seconds for CPU and around 7 seconds for memory due to parallel processing of input sequence, unlike \ac{LSTM}. The performance of \ac{SFF} and \ac{DeepAR} are almost similar in both Figures. The \ac{LSTM} uses the \ac{CPU} for a longer duration due to the sequential input processing.

\subsection{Real-World Applications}


In the domain of \ac{O-RAN} architecture enhanced by cloud-native applications, probabilistic forecasting plays a pivotal role in transforming network management. \textbf{Dynamic spectrum sharing} benefits from predictive modeling, allowing for intelligent load balancing and efficient capacity planning, thus ensuring optimal utilization of resources across diverse network segments. Additionally, \textbf{predictive maintenance} emerges as a key application, where probabilistic forecasting anticipates equipment failures, leading to improved reliability and reduced downtime. The integration of probabilistic forecasting also extends to \textbf{energy efficiency}, enabling the prediction of energy consumption patterns based on network conditions and contributing to sustainable and green networking practices. Moreover, in the \textbf{security} domain, advanced forecasting aids in proactively identifying and mitigating potential security threats, boosting the robustness of O-RAN ecosystems against evolving risks. These applications collectively showcase the transformative potential of probabilistic forecasting in O-RAN through cloud-native xApps and \ac{rApp}s, driving efficiency, reliability, sustainability, and security.





\section{Conclusion}

This paper highlights the challenges and requirements of integrating \ac{rApp} into the cloud-native \ac{O-RAN} for efficient resource allocation to meet the dynamic and complicated network resource requirements. Probabilistic forecasting of resource demand using estimators such as \ac{DeepAR} and Transformer is beneficial for service providers to make more informed and reliable decisions about \ac{PRB} demands in future time periods. The resource provisioning \ac{rApp}, which consists of a monitoring system, analytical engine, a decision engine, and an actuator, is containerized to make it suitable for use in \ac{O-RAN}. This paper then addresses the integration of containerized solutions into the O-RAN architecture for use cases related to real-world applications. The effect of data length on metrics like \ac{MSE}, training time, and prediction time is addressed for different estimators, including the deterministic \ac{LSTM} model. \ac{LSTM} performs the worst due to the sequential processing and storage of input information. Finally, the working of estimators is also compared in terms of CPU and memory usage with a data length of 20 weeks. Although \ac{SFF} performs the best with the least run time, the \ac{DeepAR} and Transformer exhibit moderate usage of resources, low \ac{MSE}, and balanced computational demands due to parallel processing of input information, self-attention mechanism, capability to capture temporal dependencies and resistance to data non-stationarity.

\section{Acknowledgements}

This work has been supported by SEMANTIC project, funded by the European Union’s Horizon 2020 research and innovation program under the Marie Skłodowska-Curie grant (agreement No 861165), the Horizon Europe project VERGE (ID: 101096034), the Spanish projects FREE6G-RadEdge (TSI-063000-2021-121) and FREE6G-RegEdge (TSI-063000-2021-144) funded by MINECO through the “NextGenerationEU” program, and the Spanish project ORIGIN (PID2020-113832RB-C22) funded MICCIN.

\bibliographystyle{IEEEtran}
\bibliography{biblio}

\begin{thebibliography}{10}
\providecommand{\url}[1]{#1}
\csname url@samestyle\endcsname
\providecommand{\newblock}{\relax}
\providecommand{\bibinfo}[2]{#2}
\providecommand{\BIBentrySTDinterwordspacing}{\spaceskip=0pt\relax}
\providecommand{\BIBentryALTinterwordstretchfactor}{4}
\providecommand{\BIBentryALTinterwordspacing}{\spaceskip=\fontdimen2\font plus
\BIBentryALTinterwordstretchfactor\fontdimen3\font minus \fontdimen4\font\relax}
\providecommand{\BIBforeignlanguage}[2]{{%
\expandafter\ifx\csname l@#1\endcsname\relax
\typeout{** WARNING: IEEEtran.bst: No hyphenation pattern has been}%
\typeout{** loaded for the language `#1'. Using the pattern for}%
\typeout{** the default language instead.}%
\else
\language=\csname l@#1\endcsname
\fi
#2}}
\providecommand{\BIBdecl}{\relax}
\BIBdecl

\bibitem{polese2023understanding}
M.~Polese \emph{et~al.}, ``Understanding o-ran: Architecture, interfaces, algorithms, security, and research challenges,'' \emph{IEEE Communications Surveys \& Tutorials}, 2023.

\bibitem{song2023micro}
P.~Song, H.~Peng, and X.~Zhang, ``A micro-service approach to cloud native ran for 5g and beyond,'' \emph{IEEE Access}, vol.~11, pp. 130\,257--130\,271, 2023.

\bibitem{KAZEMIFARD2021107809}
\BIBentryALTinterwordspacing
N.~Kazemifard and V.~Shah-Mansouri, ``Minimum delay function placement and resource allocation for open ran (o-ran) 5g networks,'' \emph{Computer Networks}, vol. 188, p. 107809, 2021. [Online]. Available: \url{https://www.sciencedirect.com/science/article/pii/S1389128621000037}
\BIBentrySTDinterwordspacing

\bibitem{article}
A.~Perveen, R.~Abozariba, M.~Patwary, and A.~Aneiba, ``Dynamic traffic forecasting and fuzzy-based optimized admission control in federated 5g-open ran networks,'' \emph{Neural Computing and Applications}, vol.~35, 06 2021.

\bibitem{xu2021survey}
Y.~Xu, G.~Gui, H.~Gacanin, and F.~Adachi, ``A survey on resource allocation for 5g heterogeneous networks: Current research, future trends, and challenges,'' \emph{IEEE Communications Surveys \& Tutorials}, vol.~23, no.~2, pp. 668--695, 2021.

\bibitem{sharma2021resource}
N.~Sharma and K.~Kumar, ``Resource allocation trends for ultra dense networks in 5g and beyond networks: A classification and comprehensive survey,'' \emph{Physical Communication}, vol.~48, p. 101415, 2021.

\bibitem{9690624}
W.~Jiang and et~al., ``Probabilistic-forecasting-based admission control for network slicing in software-defined networks,'' \emph{IEEE Internet of Things Journal}, vol.~9, no.~15, pp. 14\,030--14\,047, 2022.

\bibitem{articlemdpi}
H.~MQ and et~al., ``Recent advances in machine learning for network automation in the o-ran,'' \emph{MDPI Sensors. 2023; 23(21):8792}, 2023.

\bibitem{10266607}
V.~Kasuluru, L.~Blanco, and E.~Zeydan, ``On the use of probabilistic forecasting for network analysis in open ran,'' in \emph{2023 IEEE International Mediterranean Conference on Communications and Networking (MeditCom)}, 2023, pp. 258--263.

\bibitem{liu2022cloud}
H.~Liu, J.~Zong, Q.~Wang, Y.~Liu, and F.~Yang, ``Cloud native based intelligent ran architecture towards 6g programmable networking,'' in \emph{2022 7th International Conference on Computer and Communication Systems (ICCCS)}.\hskip 1em plus 0.5em minus 0.4em\relax IEEE, 2022, pp. 623--627.

\bibitem{mohammadi2023athena}
A.~Mohammadi and N.~Nikaein, ``Athena: An intelligent multi-x cloud native network operator,'' \emph{IEEE Journal on Selected Areas in Communications}, 2023.

\bibitem{SALINAS20201181}
\BIBentryALTinterwordspacing
D.~Salinas, V.~Flunkert, J.~Gasthaus, and T.~Januschowski, ``Deepar: Probabilistic forecasting with autoregressive recurrent networks,'' \emph{International Journal of Forecasting}, vol.~36, no.~3, pp. 1181--1191, 2020. [Online]. Available: \url{https://www.sciencedirect.com/science/article/pii/S0169207019301888}
\BIBentrySTDinterwordspacing

\bibitem{NIPS2017_3f5ee243}
A.~Vaswani, N.~Shazeer, N.~Parmar, J.~Uszkoreit, L.~Jones, A.~N. Gomez, {\L}.~Kaiser, and I.~Polosukhin, ``Attention is all you need,'' \emph{Advances in neural information processing systems}, vol.~30, 2017.

\bibitem{alexandrov2018gluonts}
A.~Alexandrov, S.~De, V.~Dzyuba, J.~Li, and J.~Tejedor, ``Gluonts: probabilistic and neural time series modeling in python,'' \emph{arXiv preprint arXiv:1906.05264}, 2019.

\bibitem{9933014}
F.~Rezazadeh \emph{et~al.}, ``On the specialization of fdrl agents for scalable and distributed 6g ran slicing orchestration,'' \emph{IEEE Transactions on Vehicular Technology}, vol.~72, no.~3, pp. 3473--3487, 2023.

\end{thebibliography}

\end{document}